\documentclass[aps,showpacs,nofootinbib,superscriptaddress]{revtex4}
\usepackage[english]{babel}
\usepackage[latin1]{inputenc}
\usepackage{amsmath}
\usepackage{amssymb}
\usepackage{graphicx}
\usepackage{epsfig}

\begin{document}
\title{Local Fermi gas in inclusive muon capture from nuclei}
\author{J.E. Amaro}
\affiliation{Departamento de F\'{\i}sica At\'omica, Molecular y Nuclear, 
          Universidad de Granada,
          Granada 18071, Spain}
\author{J. Nieves}
\affiliation{Departamento de F\'{\i}sica At\'omica, Molecular y Nuclear, 
          Universidad de Granada,
          Granada 18071, Spain}
\author{M. Valverde}
\affiliation{Departamento de F\'{\i}sica At\'omica, Molecular y Nuclear, 
          Universidad de Granada,
          Granada 18071, Spain}
\author{C. Maieron}
\affiliation{INFN, Sezione di Catania,
              Via Santa Sofia 64, 
              95123 Catania, Italy}

\begin{abstract}
  We compare local Fermi gas and shell model in muon capture in 
  nuclei in order to estimate the effect of finite nuclear size in
  low energy weak reactions.
\end{abstract}

\maketitle

\section{Introduction}

In this paper we study the importance of nuclear finite size effects
in inclusive muon capture reactions. 
The inclusive muon capture process in nuclei 
\begin{equation}  \label{decay}
  \mu^- + ^A_Z X \to X + \nu_\mu 
\end{equation}
is very similar to neutrino scattering off nuclei 
\begin{equation}  \label{nuscat}
  \nu_\mu + ^A_Z X \to X + \mu^-
\end{equation}
and is experimentally more accesible so it serves as a benchmark for 
testing theoretical models of the latter process. 

The motivation for this investigation comes from 
the results of a model (published in \cite{Nie04}),  
which describes rather well inclusive
$^{12}$C$(\nu_\mu,\mu^-)X$ and $^{12}$C$(\nu_e,e^-)X$ cross sections
near threshold and inclusive muon capture by nuclei.
This approach, which is an extension of the quasi--elastic (QE)
inclusive electron scattering model of \cite{Gil97}, is based on a
Local Fermi Gas (LFG), where the simplicity of the model makes it
possible to include a great variety of effects into the reaction dynamics.
The goal of this paper is to investigate whether finite nucleus
effects that are not adressed by a LFG can affect significantly the results of 
the model in \cite{Nie04}.

There already exist microscopic calculations of neutrino--nucleus
reactions and muon capture that treat correctly the finite size of the
system.  However there are some dynamical issues that are implemented
in a different fashion in a LFG model so a direct comparison of these
models is not possible and we can not extract the effect of the inclusion 
of nuclear structure details.  For this reason, we perform a
comparison of the LFG model with a extreme shell model
(SM), that is single particle states in a Woods-Saxon (WS) potential, 
where the finite size effects can be easily recognized.

In the second section we will introduce the model of \cite{Nie04} as applied to
inclusive nuclear muon capture and give some numerical results to be
compared with experiment. Then we will introduce the extreme SM
and a simplified LFG model. In the fourth section we will show the comparison 
between the two models. We will finish with some conclusions.

More details on this issue can be found in \cite{Ama05}. A more comprehensive
analysis of the uncertainties in the model of \cite{Nie04} 
can be found in \cite{Val06}.

\section{Inclusive Muon Capture in  Nuclei}
The evaluation of the decay width for inclusive muon capture in finite nuclei 
proceeds in two steps. 
In the first one we evaluate the spin averaged decay width for
a muon at rest in a Fermi sea of protons and neutrons 
$\hat\Gamma\left(\rho_p,\rho_n\right)$
with $N\neq Z$. In the second step, we use the LFG approximation to go to
finite nuclei and evaluate
\begin{equation}
\Gamma = \int d^3 {\bf r} |\phi_{\rm 1s}({\bf r})|^2
\hat\Gamma\left(\rho_p(r),\rho_n(r)\right ) \label{eq:imclda}
\end{equation}
where $\phi_{\rm 1s} (\vec{r}\,)$ is the muon wave function in the
$1s$ state from where the capture takes place. It has been obtained by
solving the Schr\"odinger equation with a Coulomb interaction taking
into account the finite size of the nucleus and vacuum
polarization. This approximation assumes a zero range of the
interaction, that becomes highly accurate as long as the $\vec{q}$
dependence of the interaction is extremely weak for the $\mu-$atom
decay process.

The spin averaged muon decay width is related to the imaginary
part of the self-energy of a muon at rest in the medium.
Further details and analytical expressions can be found in the 
Appendix of \cite{Nie04}.  

For kinematical reasons only the QE part of the $W^-$ self--energy
contributes to the muon decay. 
Thus, both the muon decay in the medium (Eq.~\ref{decay}) and the electroweak
inclusive nuclear reactions (Eq.~\ref{nuscat}) 
in the QE regime are sensitive to the same physical features.
We can apply the same nuclear physics corrections to the above model as 
in neutrino scattering,
that is Pauli blocking, RPA corrections and corrections to the 
energy balance, see \cite{Nie04}.
The $1s$ muon binding energy, $B_\mu^{1s}>0$, is taken
into account by replacing $m_\mu \to \hat m_\mu= m_\mu - B_\mu^{1s}$.

In muon capture only very small nuclear excitation energies are explored, 
0--25MeV, so the kinematical regime of the muon capture process is the 
worst possible for a LFG model of the nucleus. 
Nevertheless, the predictions of this model are in fairly good
agreement with the experimental results as can be seen in 
Table~\ref{tab:capres}.
\begin{table}\small{
\begin{center} 
\begin{tabular}{ccc|cc}\hline
  & Pauli+$\overline{Q}$ $[10^4\,s^{-1}]$ & RPA $[10^4\, s^{-1}]$ &
  Exp $[10^4\, s^{-1}]$ & $\delta_{\mbox{rel}}\Gamma$ \\\hline
  $^{12}$C & 5.42 & 3.21 & $3.78\pm 0.03$ & \phantom{$-$}0.15 \\ 
  $^{16}$O & 17.56 & 10.41 & $10.24\pm 0.06$ & $-0.02$ \\ 
  $^{18}$O & 11.94 & 7.77 & $8.80\pm 0.15$ & \phantom{$-$}0.12 \\
  $^{23}$Na &58.38 & 35.03 & $37.73\pm 0.14$ & \phantom{$-$}0.07 \\
  $^{40}$Ca &465.5 &257.9 &$252.5\pm 0.6 $ & $-0.02$ \\
  $^{44}$Ca &318 &189 & $179 \pm 4 $ & $-0.06$ \\
  $^{75}$As &1148 & 679 & 609$\pm 4$ & $-0.11$ \\ 
  $^{112}$Cd &1825 & 1078 & 1061$\pm 9 $ & $-0.02$ \\ 
  $^{208}$Pb & 1939 & 1310 & 1311$\pm 8 $ & \phantom{$-$}$0.00$ \\ \hline 
\end{tabular}
\end{center} 
\caption{\footnotesize Experimental and theoretical total muon
  capture widths for different nuclei. 
  We quote two different theoretical
  results: i) Pauli+$\overline{Q}$: obtained without including RPA
  correlations, but taking into account the value of
  $\overline{Q}$; ii)RPA: the full calculation, including all nuclear effects.
  Experimental data (Exp) are a weighted average:
  $\overline{\Gamma}/\sigma^2 = \sum_i \Gamma_i/\sigma_i^2$, with
  $1/\sigma^2 = \sum_i 1/\sigma_i^2$ of the results cited in \cite{Su87}.
  Finally, in the last column we show the relative discrepancies
  existing between the theoretical predictions given in the third column and 
  the experimental data of the fourth column. 
($\delta_{\mbox{rel}}\Gamma=\left(\Gamma^{\rm Exp}-\Gamma^{\rm Th}\right )
/\Gamma^{\rm Exp} $)}
\label{tab:capres}} 
\end{table}

\section{Comparison of non-correlated models}

In order to simplify the calculations in the comparison between the two models 
we make a static
approximation and expand the single nucleon weak current $J^\mu$ in the nucleon
momentum  keeping terms up to order zero.

In the SM we have to deal with an $S$ matrix element of the kind
\begin{equation}
  S_{fi} = -2\pi i\delta(E_f-E_i-\omega)\frac{G}{\sqrt{2}}{\ell}^\mu
  \langle f|\tilde{J}_\mu(-{\bf k}')|i\rangle
\end{equation}
where the weak current is modified to include the muon wave function
  \begin{equation}
    \tilde{J_\mu}(-{\bf k}^\prime) =
    \int d^3{\bf r}\,{\rm e}^{-i{\bf k}^\prime\cdot{\bf r}}
    J_\mu({\bf r})\phi_{1s}({\bf r}) .
  \end{equation}
Now the states $|i\rangle$ and $\langle f|$ are nuclear states 
in a shell model.
This states are single particle excitations of a N nucleon system in a WS 
potential and can be not only in the continuum, like in a Fermi Gas, but
there can also be discrete excitations so we have now two contributions 
to the decay width.

We now have to get the nuclear wave functions as solutions of the Schrödinger 
equation for a WS potential where
the parameters of the potential are commonly  fitted  to 
the experimental energies of the valence shells or the 
charge radius. In the present case of muon capture 
we fit the experimental $Q$-value for the decay reaction.

For the LFG model we use simplified expression for the decay width,
where we have not taken into account RPA correlations, but energy balance
and Pauli blocking effects are implemented.
The only inputs remaining to be fixed are the nuclear matter densities,
that will be those provided by the wave functions of the WS potential.

Up to now we have in both LFG and SM models the same physical features
of Pauli blocking and correct energy balance, the only difference
coming from the more refined treatment of the nuclear wave function
in the SM case. 
In this way we can compare both SM and LFG models.

\section{Results}

In this section we present results for a set of closed--shell nuclei 
$^{12}$C, $^{16}$O, $^{40}$Ca and $^{208}$Pb.
Fixing the experimental $Q$-value only makes one condition for fixing 
the several parameters of the WS potential, so
wherever possible we set the remaining  parameters of
the potential to values similar to those used in the literature.
In our calculation we use different sets of 
parameters, denoted WS1, WS2 and WS3.
In order to compare with the LFG, we use 
as input the proton and neutron densities obtained in the corresponding 
shell model. The values of the different parameters sets can be seen
in \cite{Ama05}.

In Table~\ref{tableIV} we show results for the integrated inclusive widths for
the four nuclei.
\begin{table} 
\begin{center}
\begin{tabular}{cccccr}
         &     & discrete  &  total  & LFG     &     \%   \\ \hline\hline
$^{12}$C & WS1 &   0.3115  & 0.4406  & 0.4548  &    3.2   \\ 
         & WS2 &   0.3179  & 0.4289  & 0.4360  &    1.7   \\ 
         & WS3 &   0.2746  & 0.5510  & 0.4732  & $-14.1$  \\ \hline
$^{16}$O & WS1 &   1.124   & 1.267   & 1.346   &    6.2   \\ 
         & WS2 &   0.584   & 1.107   & 1.378   &   24.4   \\ 
         & WS3 &   1.143   & 1.316   & 1.373   &    4.3   \\ \hline
$^{40}$Ca& WS1 &  27.72    & 34.87   & 34.81   &  $-0.1$  \\
         & WS2 &  26.34    & 31.70   & 33.07   &    4.3   \\
         & WS3 &  24.91    & 30.64   & 33.19   &    8.3   \\\hline
$^{208}$Pb&WS1 &  128.5    & 191.0   & 187.27  &  $-1.9$  \\
          &WS2 &  159.6    & 243.4   & 213.64  &  $-12.2$ \\ \hline\hline
\end{tabular}
\caption{\label{tableIV} Integrated width in units of $10^{5}$s$^{-1}$ 
for the different nuclei and Woods-Saxon potentials, compared with the 
LFG results using the corresponding charge densities. 
The discrete contribution of the shell model is shown in the first column.
The column labeled \% gives the relative diference between WS and LFG results.}
\end{center}
\end{table}
We can see that, in the case of WS1 and WS2, the LFG and WS results for
$^{12}$C are 
quite similar, differing only in $\sim 2$--3\%. In the case of WS3 the
differences are larger, around 14\%. 
\begin{figure}
\begin{center}
\begin{tabular}{cc}
\includegraphics[scale=0.5]{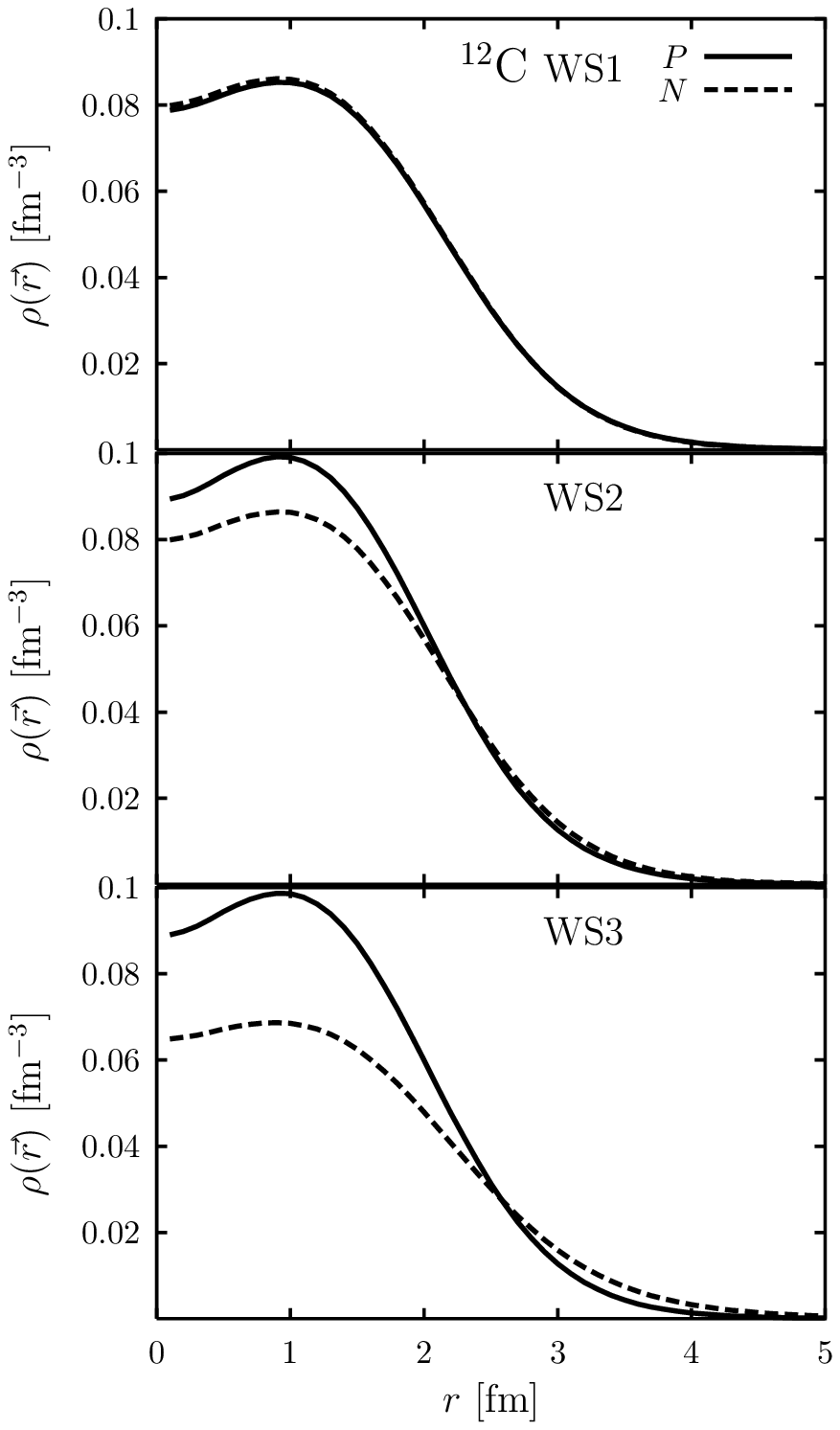}&
\includegraphics[scale=0.5]{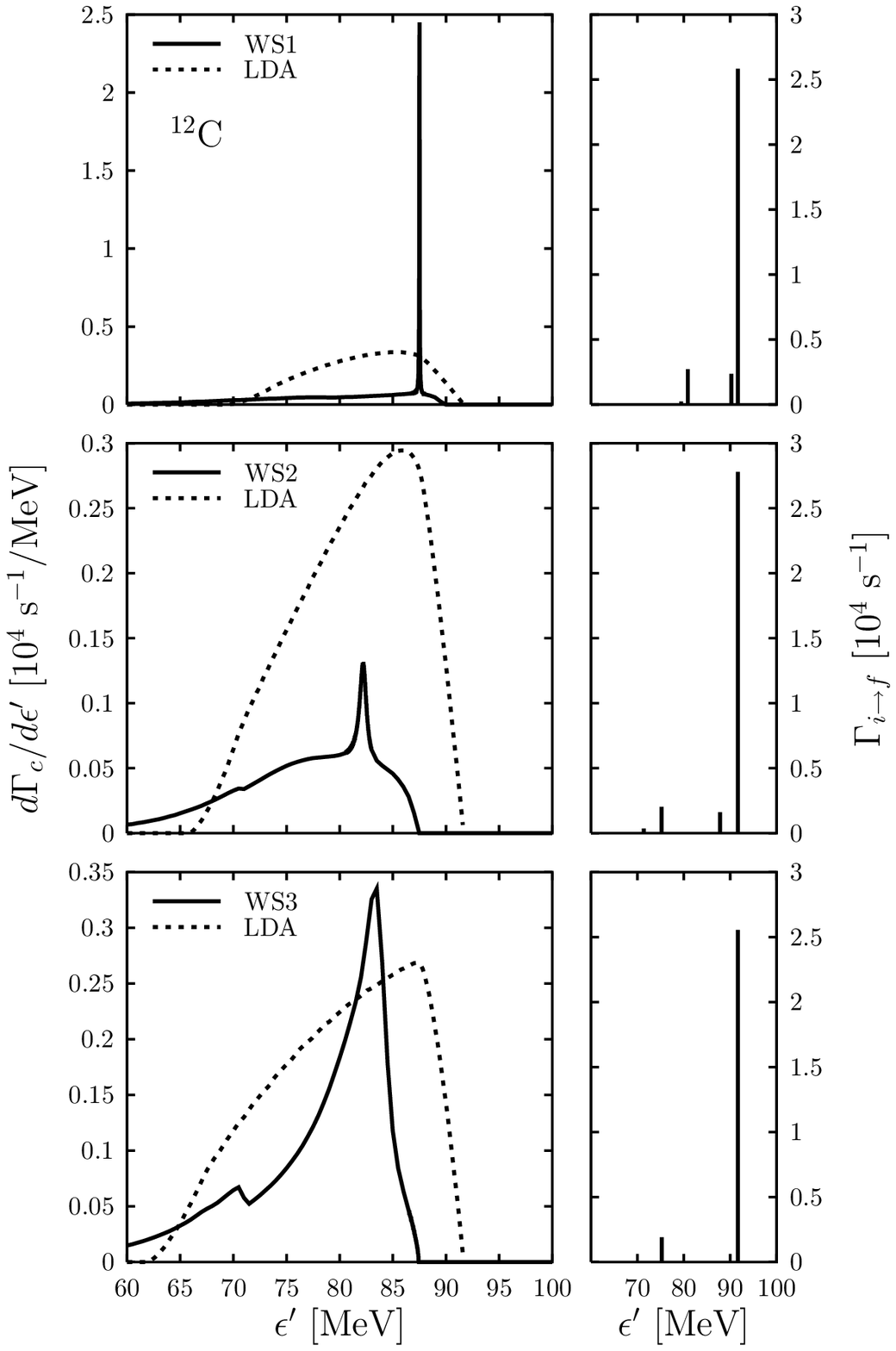}
\end{tabular}
\end{center}
\caption{The left figure shows proton and neutron densities of $^{12}$C,
 for the several WS potentials used in this work.
 The right figure shows the differential SM width of $^{12}$C to the continuum 
 (left panels) compared to the LFG, and partial widths contributions to the 
 discrete states (right panels), as a function of the neutrino energy.}
\label{fig:cdens}
\end{figure}

This can be understood in terms of the values of the parametrization WS3.
For more attractive potentials
the nucleus becomes more dense in the interior.  For this reason, the WS3
neutron density turns out to be the smallest one, while the proton density is
around 3/2 the neutron one. Therefore a proton near the Fermi surface can
decay to a neutron above the neutron Fermi surface with an energy decrement.
This is an unrealistic situation, since precisely in this case the neutrons
are less bound than protons in the SM, and therefore lie at higher energies.
Hence the LFG results are worse for very
different neutron and proton densities. 
Another argument to disregard this case is the well known property of closed
(sub--)shell nuclei such as $^{12}$C, 
for which the neutron and proton densities should be similar.

In Fig.~\ref{fig:cdens} we compare the SM results for the differential width 
to the continuum
with the LFG distribution for the different WS parameters (left panels). The
shapes of both distributions are completely different.
The partial widths to the discrete
states are shown in the right panels of Fig.~\ref{fig:cdens}.  
Considering these differences
in shape between the LFG and the SM, it is a very notable result that the
integrated widths (adding the discrete states) take similar values  in both
models as was shown in Table~\ref{tableIV}. 

\begin{figure}
\begin{center}
\begin{tabular}{cc}
\includegraphics[scale=0.5]{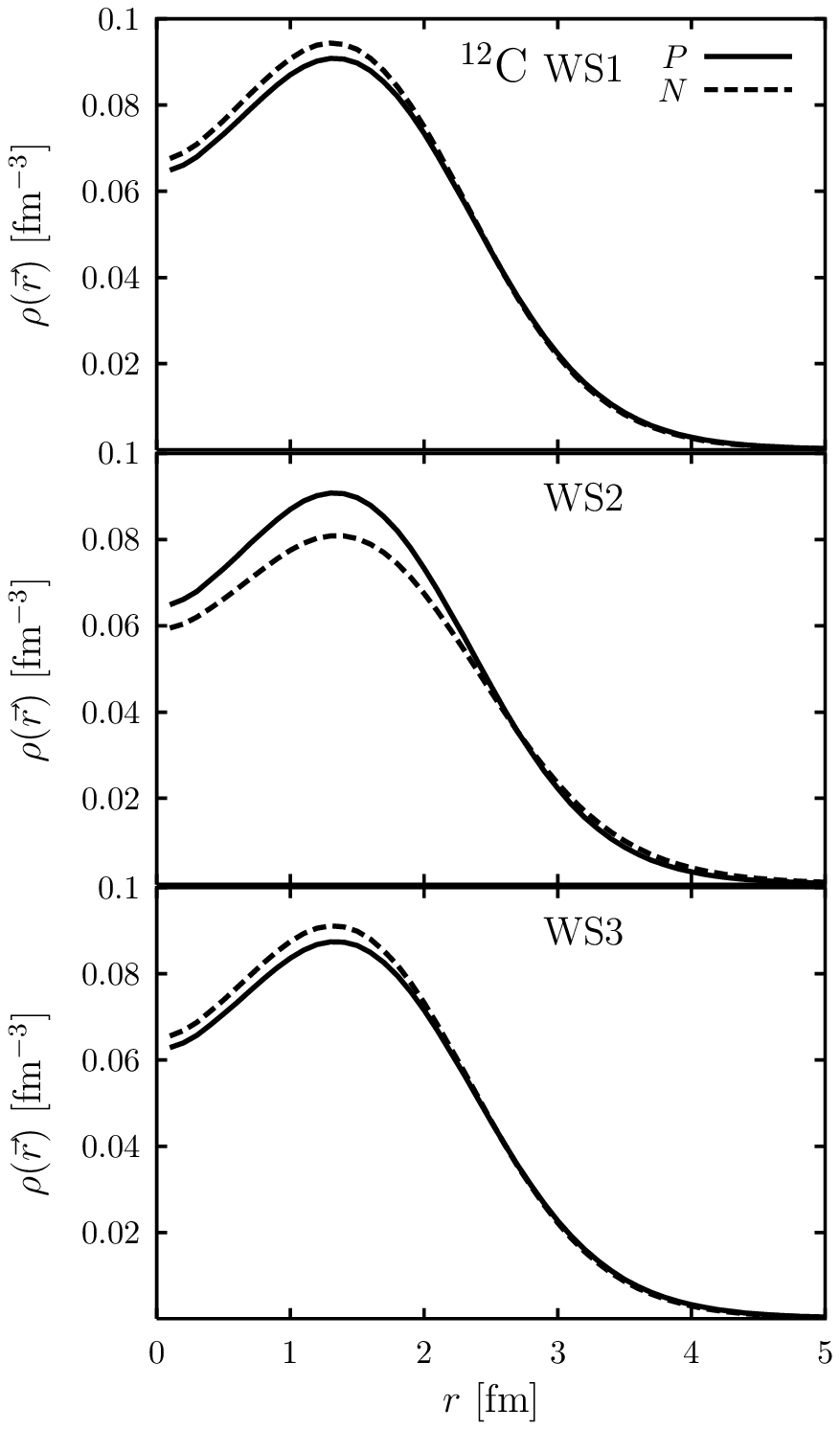} &
\includegraphics[scale=0.5]{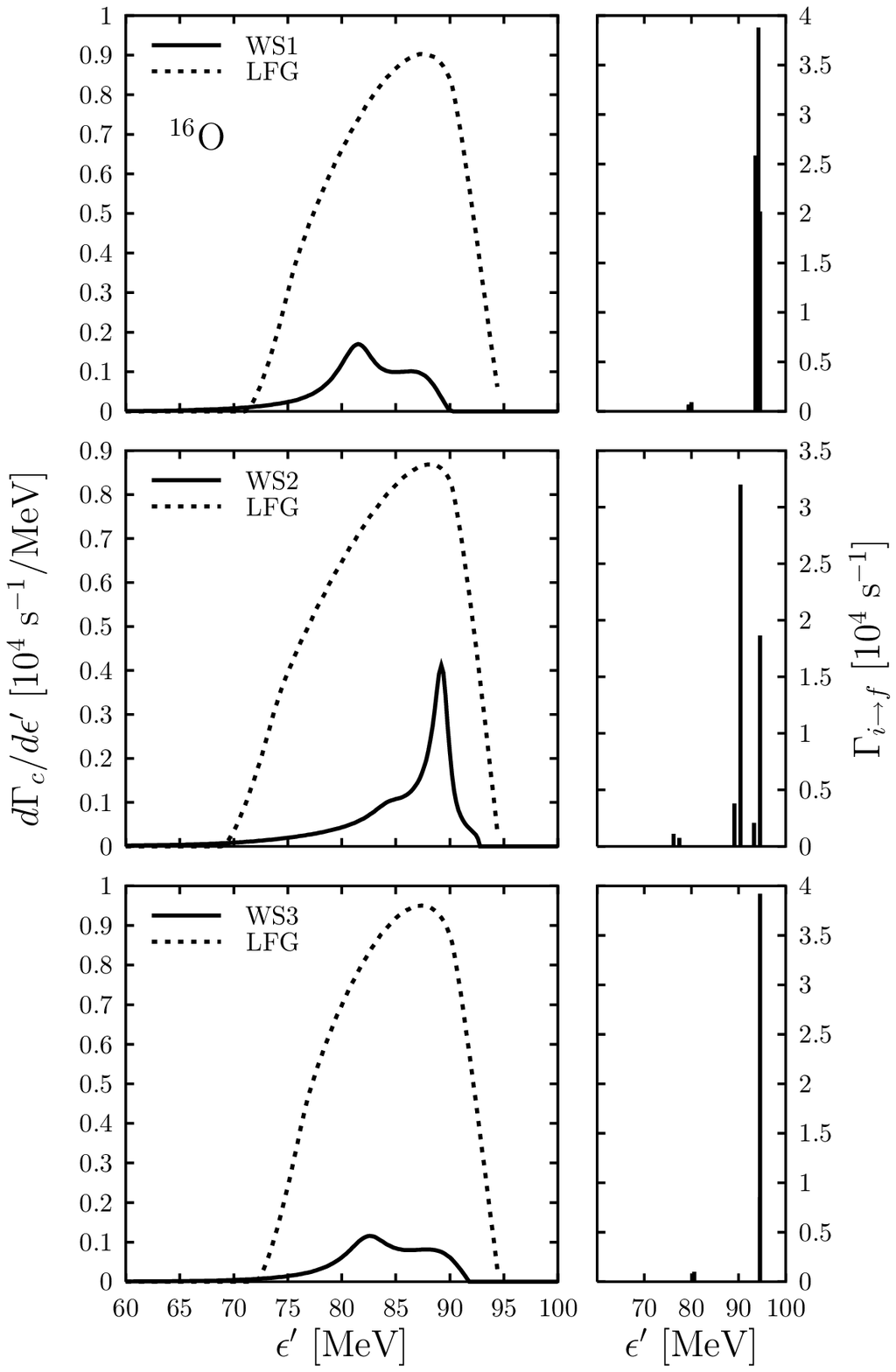}
\end{tabular}
\end{center}
\caption{The same as Fig.~\ref{fig:cdens} for 
$^{16}$O.}
\label{fig:odens}
\end{figure}

In the case of $^{16}$O the integrated widths computed in the LFG are also
very close, $\sim 4$--6\%, to the SM results with the potentials WS1 and WS3
(see Table~\ref{tableIV}).  The worse results are obtained for the WS2
parameterization; the corresponding width is 24\% of the SM one.  This can
also been understood in terms of what was said for the case of $^{12}$C above,
by looking at the $^{16}$O densities shown in Fig.~\ref{fig:odens}.

The LFG results improve when the mass of the nucleus increases as in the case 
of the nucleus $^{40}$Ca. In fact, from Table~\ref{tableIV} we see that for
this nucleus the LFG integrated width is lower than 8\% for all cases.
This improvement was expected because the Fermi gas description of the 
nucleus should work better for heavier nuclei.
\begin{figure}
\begin{center}
\begin{tabular}{cc}
\includegraphics[scale=0.5]{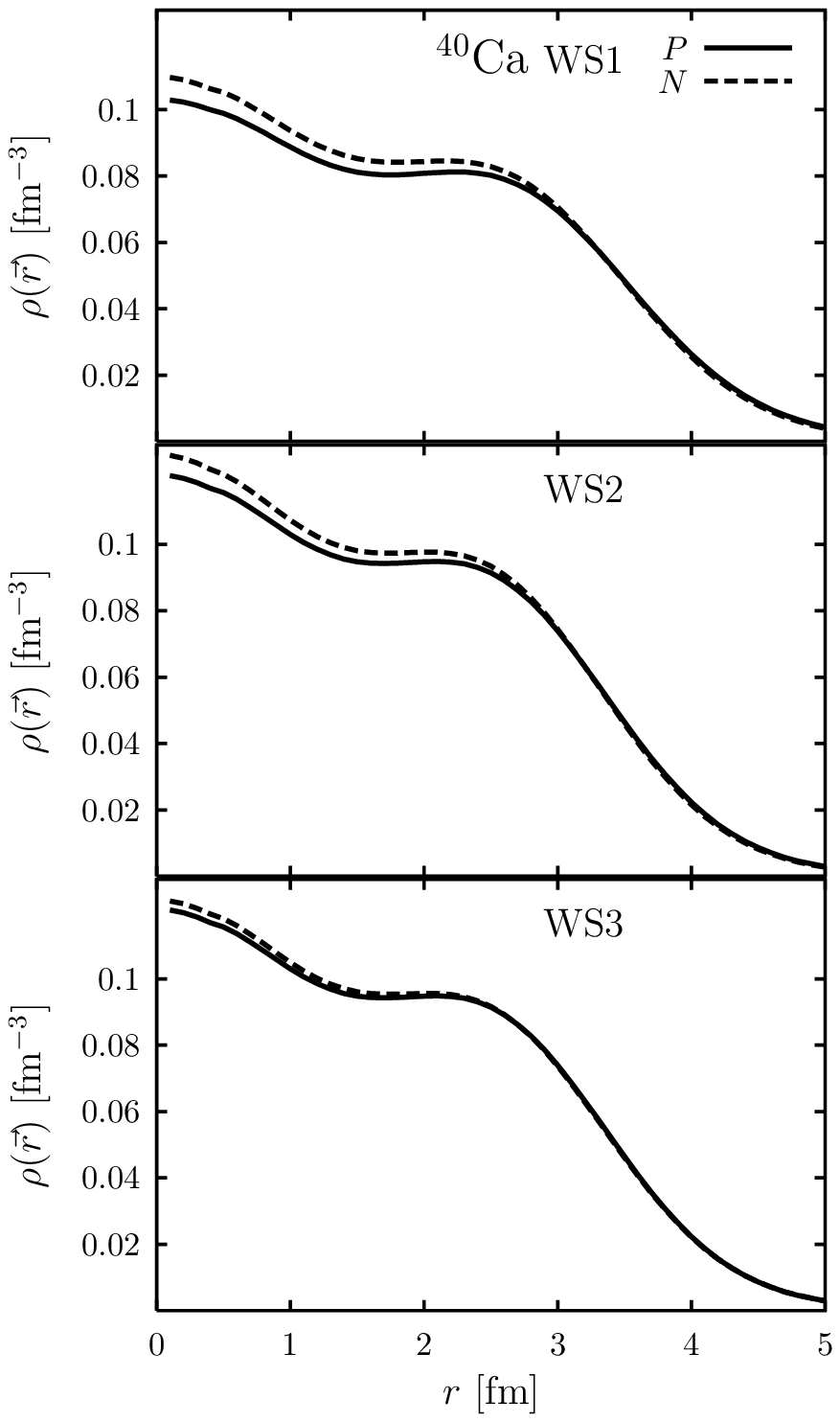} &
\includegraphics[scale=0.5]{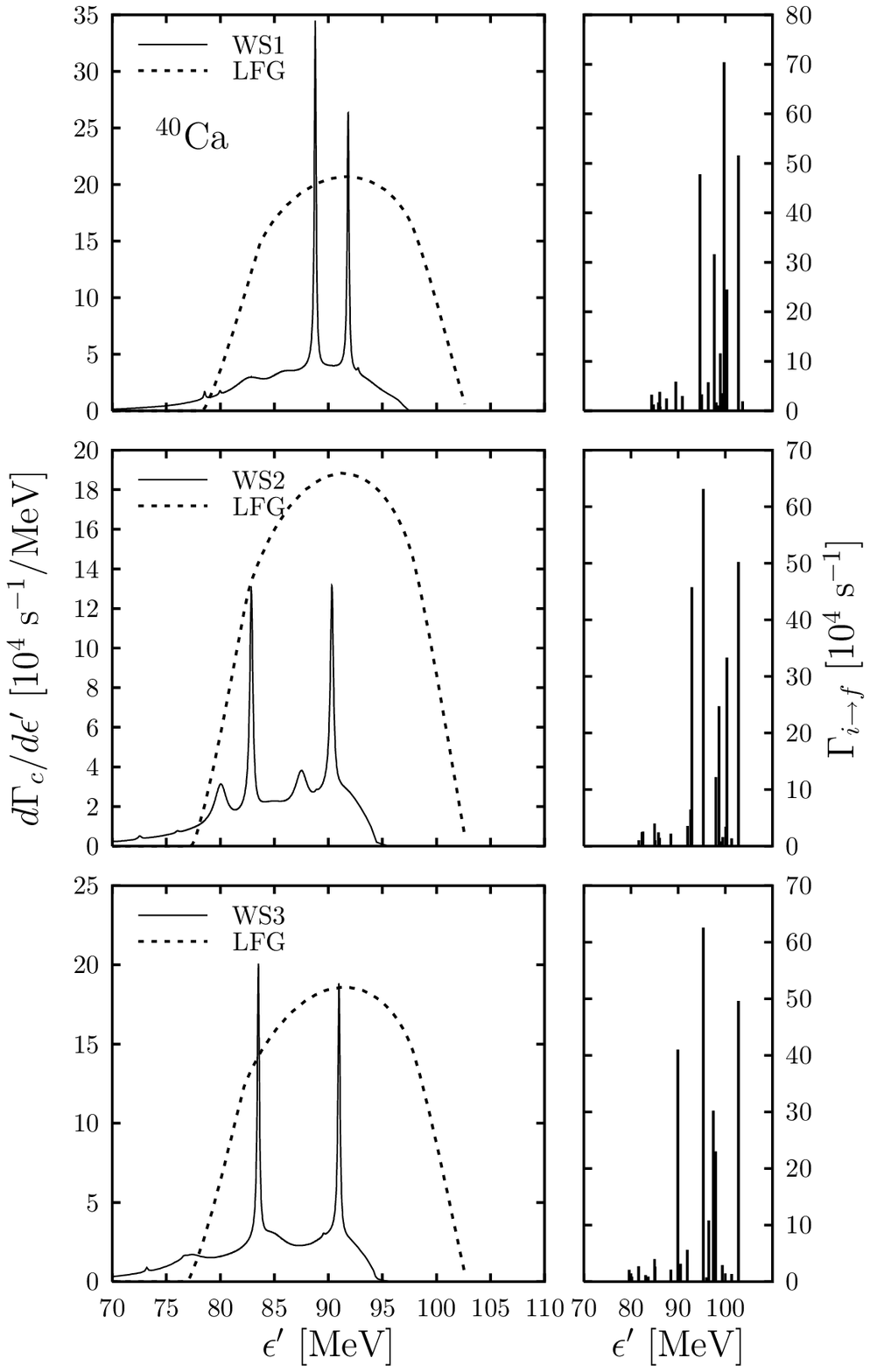}
\end{tabular}
\end{center}
\caption{The same as Fig.~\ref{fig:cdens} for $^{40}$Ca.}
\label{fig:cadens}
\end{figure}

For the closed-shell heavy nucleus $^{208}$Pb we present
in Table~\ref{tableIV} integrated widths only for two sets of
potential parameters, WS1 and WS2. 
In both cases the LFG results are close to the SM ones.
\begin{figure}[t]
\begin{center}
\begin{tabular}{cc}
  \includegraphics[scale=0.5]{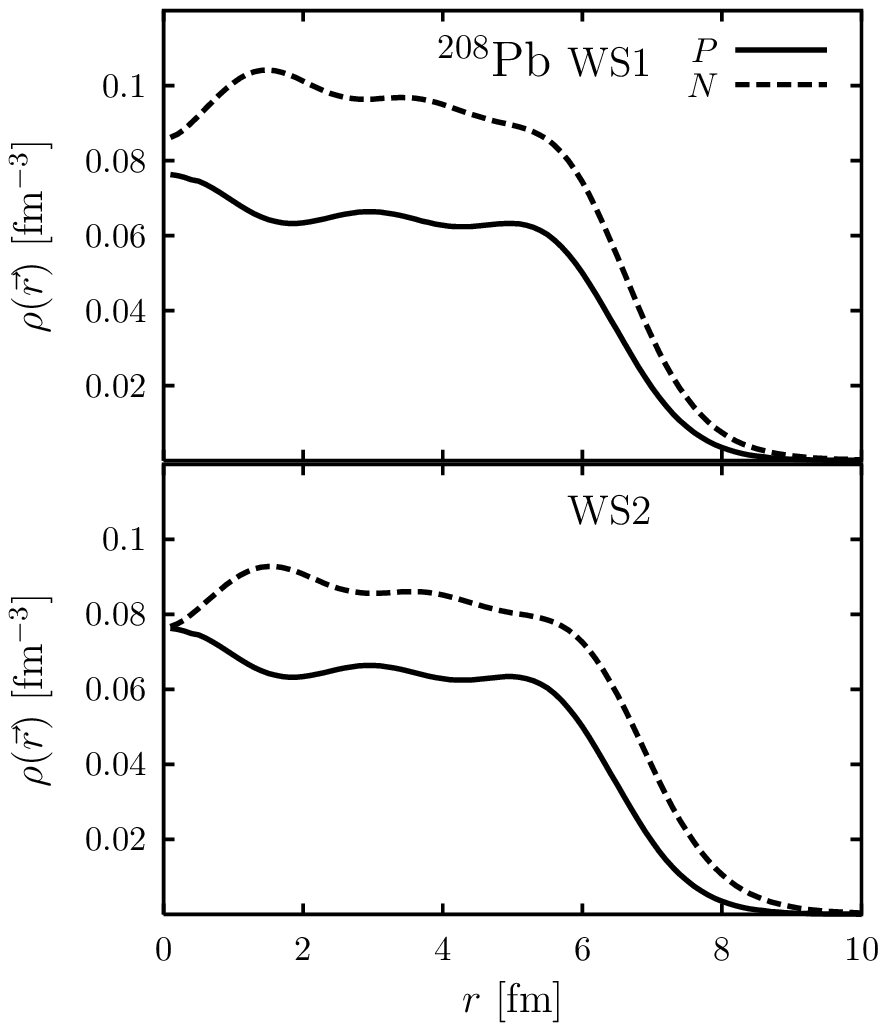}&
  \includegraphics[scale=0.5]{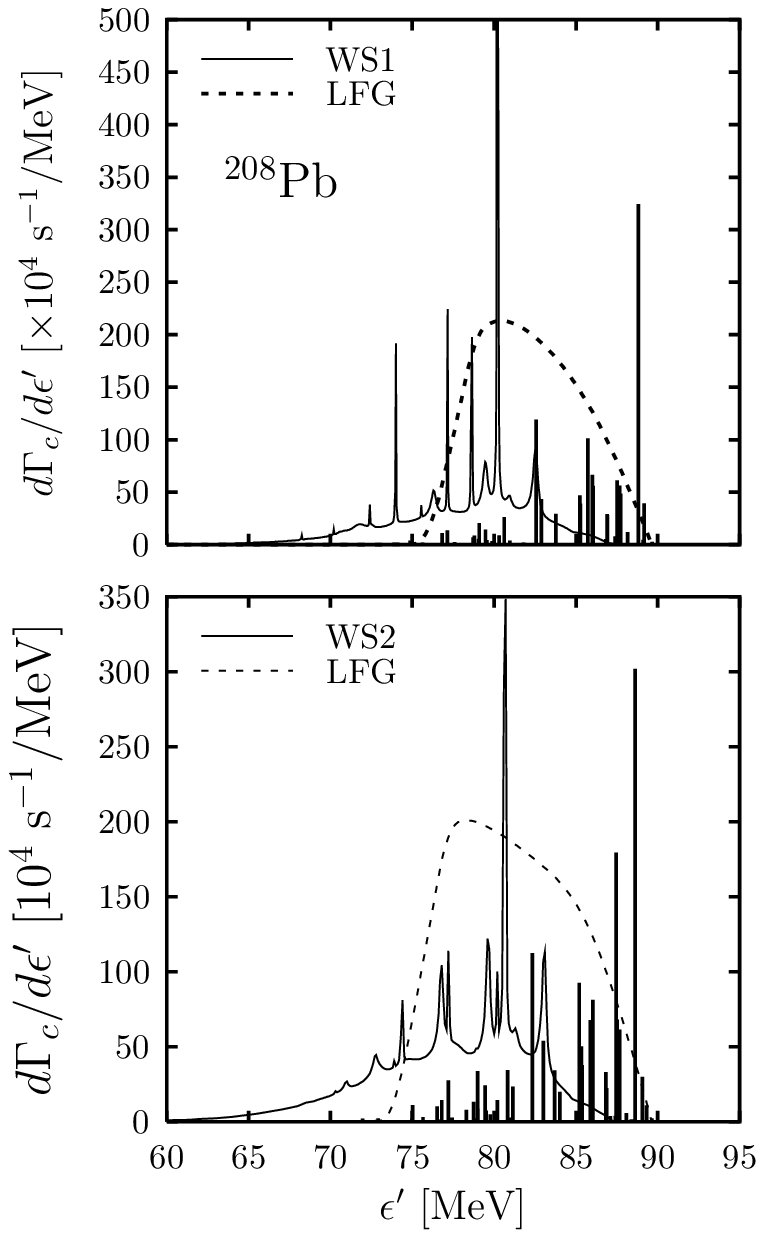}
\end{tabular}
\end{center}
\caption{The same as Fig.~\ref{fig:cdens} for $^{208}$Pb.}
\label{fig:pbdens}
\end{figure}

\section{Conclusions}

In this paper we have estimated the magnitude of the finite nucleus effects on
inclusive muon capture, aiming at quantifying the uncertainty of the LFG
results of \cite{Nie04}.

We have focused on a simple shell model without nuclear
correlations, but that contains the relevant information about the finite
nuclear structure, and we have compared it with the uncorrelated LFG using the
same input. As expected, the neutrino spectrum is very
different in the two models, in particular the LFG cannot account for the
resonances and discrete states.  However, in the case of the lighter nuclei,
$^{12}$C and $^{16}$O, the SM and LFG results for the integrated width are
close ---within 3--6\%--- for WS parameters with similar neutron and proton
densities.  For the medium
and heavy nuclei, $^{40}$Ca and $^{208}$Pb, the integrated widths are always
very close, within 1--7\%.  The final neutrino spectra of the LFG become more
similar to the SM, including the discrete part, for heavier nuclei.

\begin{acknowledgments}
This work was supported by DGI and FEDER funds, contracts
BFM2005-00810 and BFM2003-00856 and by the Junta de Andaluc\'\i a (FQM-225).
\end{acknowledgments}


\end{document}